\documentclass[fleqn,12pt,twoside]{article}
\usepackage{espcrc1}

\usepackage{graphicx}
\usepackage[figuresright]{rotating}

\newcommand{\AmS}{{\protect\the\textfont2
  A\kern-.1667em\lower.5ex\hbox{M}\kern-.125emS}}

\title{Electromagnetic interactions in nucleus-nucleus and proton-proton collisions}

\author{J. Nystrand\address{Department of Physics and Technology, 
        University of Bergen, \\ 
        All{\'e}gaten 55, N-5007 Bergen, Norway}
}
       
\begin{document}

% typeset front matter
\maketitle

\begin{abstract}
The strong electromagnetic fields associated with ultra-relativistic 
protons and nuclei correspond to an equivalent flux of photons. At the 
future Large Hadron Collider at CERN, the corresponding photon-nucleon 
center of mass energies will be higher than at any existing accelerator. 
The experimental and theoretical aspects of particle production in 
electromagnetic interactions at hadron colliders are reviewed. 
\end{abstract}

\section{Introduction}

Traditionally, two-photon and high-energy photon-nucleon/nucleus interactions have been 
studied using electron beams. Two-photon interactions have been studied at 
e$^+$e$^-$ colliders such as PEP, Petra, and LEP. 
Photon-proton interactions have 
been studied in fixed target experiments using electron beams and at the 
electron-proton collider HERA. The interest in utilizing the strong photon flux associated 
with ultra-relativistic protons and nuclei to study these processes is, however, 
growing. See \cite{Baur:2002vr} for recent reviews. 

The main advantage of using colliding hadron and nuclear beams for studying photon 
induced interactions is the high equivalent photon energies and luminosities that 
can be achieved at 
existing and future accelerators. These include the Relativistic Heavy Ion Collider (RHIC) 
at Brookhaven National Laboratory, the Fermilab Tevatron, and the Large Hadron Collider (LHC) 
at CERN. 

The disadvantages are that it is not possible to tag the outgoing protons or nuclei, and 
that the interactions will always be restricted to low-virtuality photons. 
Experimentally, triggering on photon induced interactions is also a challenge, since most 
experiments at hadron colliders so far have been constructed for studying hadronic interactions, 
which have a very different event topology.

\section{The equivalent photon flux at hadron colliders}

In the equivalent photon approximation, the effect of the electromagnetic field of a relativistic, charged 
particle is replaced by a flux of photons. The spectrum of photon energies, $k$, may in 
principle extend up to the energy, $E$, of the projectile but is in practice only significant at a small 
fraction, $x = k/E$, of that. For an extended object, such as a proton or a nucleus, the energy spectrum 
can be calculated from \cite{Budnev:1974de}:
\begin{equation}
\label{f_x}
f(x) = \frac{dn_{\gamma}}{dx} = \frac{\alpha Z^2}{\pi} \, \frac{1 - x + 1/2 x^2}{x} 
\int_{Q_{min}^2}^{\infty} \frac{Q^2 - Q_{min}^2}{Q^4} | F(Q^2) |^2 dQ^2 \; .
\end{equation}
Here, $Q^2$ is the 4-momentum transfer from the projectile, which has a form factor $F(Q^2)$, and  
$\alpha$ is the fine structure constant. We use units in which $\hbar = c = 1$. The minimum 
momentum transfer, $Q_{min}^2$, is a function of $x$ and is to a good approximation given by 
$Q_{min}^2 = (x M_A)^2/(1-x)$, where $M_A$ is the mass of the projectile. 

\begin{figure}
\begin{center}
\includegraphics[width=12cm]{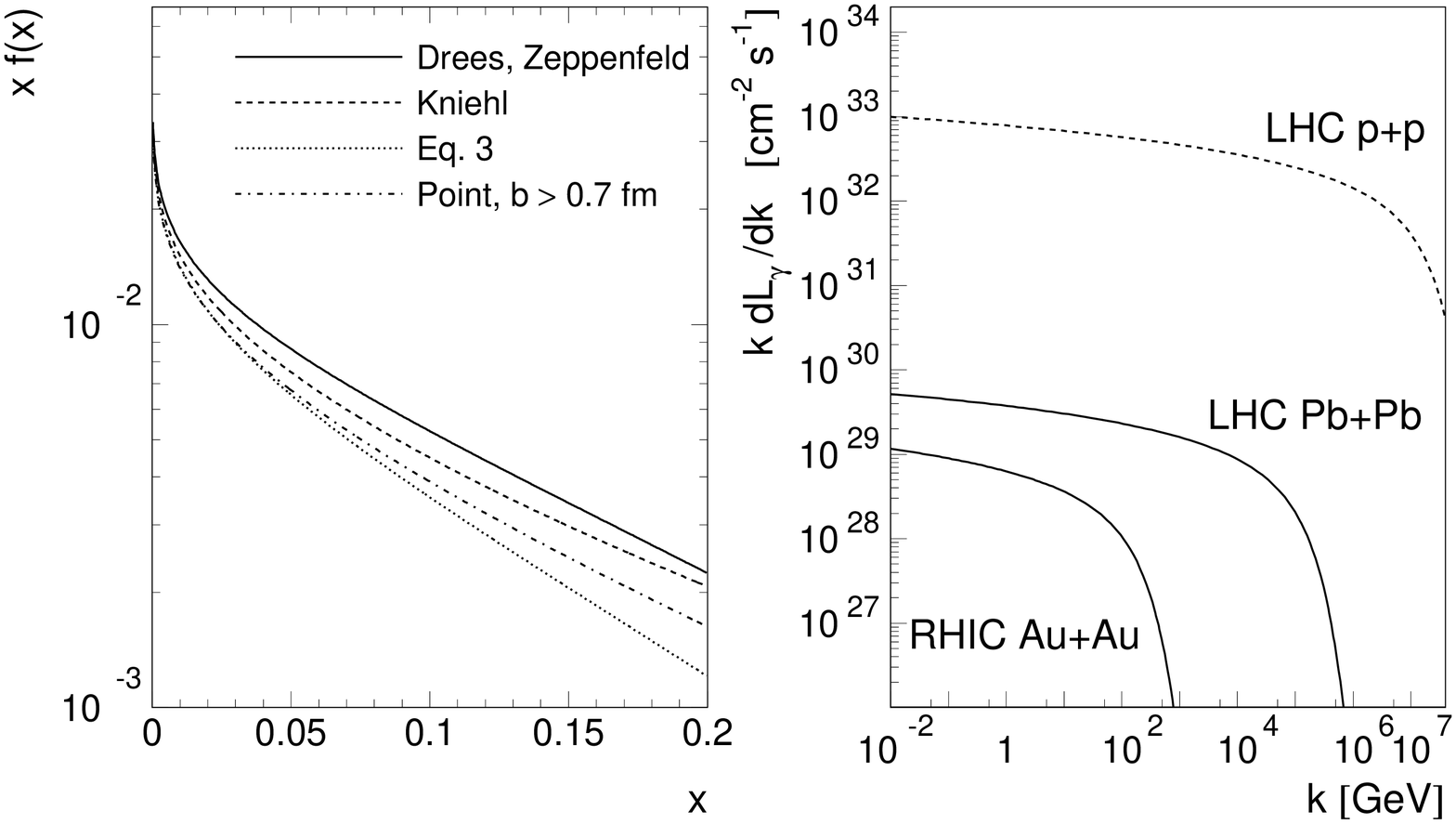}
\caption{Left: A comparison of different calculations of the equivalent photon spectrum for 
high-energy protons. $x$ is the fraction of the proton energy carried by the photon. 
Right: The equivalent photon luminosities in nucleus-nucleus and proton-proton 
collisions at RHIC and the LHC. $k$ is the photon energy in the rest frame of the target.}
\label{photon_spectrum}
\end{center}
\end{figure}

The equivalent photon spectrum of high energy protons has been calculated by Drees and 
Zeppenfeld\cite{Drees:1988pp}. 
Using the electric dipole form factor $F_E (Q^2) = 1/(1 + Q^2/0.71 GeV^2)^2$ they find 
\begin{equation}
\label{DZ}
f(x) = \frac{\alpha}{\pi} \, \frac{1 - x + 1/2 x^2}{x} \left[ \ln(A) - \frac{11}{6} +
\frac{3}{A} - \frac{3}{2 A^2} + \frac{1}{3 A^3} \right]
\end{equation}
where $A = 1 + (0.71 GeV^2)/Q_{min}^2$. This result can be derived from Eq.~\ref{f_x} 
if the second term containing $Q_{min}^2$ inside the integral is neglected, i.e. making the 
approximation $(Q^2 - Q_{min}^2)/Q^4 \approx 1/Q^2$. If this term is included, $f(x)$ is 
\begin{equation}
\label{JN}
f(x) = \frac{\alpha}{\pi} \, \frac{1 - x + 1/2 x^2}{x} 
\left[ \frac{A+3}{A-1} \ln(A) - \frac{17}{6} - \frac{4}{3A} + \frac{1}{6 A^2} \right]  \; .
\end{equation}
The two results will be compared below. The effect of including the magnetic dipole moment and the 
corresponding magnetic form factor of the proton has been 
investigated by Kniehl\cite{Kniehl:1990iv}. The final result (Eq. 3.11 of \cite{Kniehl:1990iv}) 
is too long to include here, but will be discussed further below. 

Equation~\ref{f_x} may in principle be used also for the photon spectrum of relativistic heavy ions, 
with an appropriate form factor. For a collision between two heavy ions it is, however, more appropriate to 
calculate the spectrum of equivalent photons as a function of impact parameter\cite{Cahn:1990jk,Baur:1990fx}. 
The advantage is that in this representation the contribution from 
interactions in which the ions interact hadronically can easily be excluded. 

The photon energy spectrum produced by a point particle sweeping past a target at a minimum 
impact parameter, $b_{min}$, can be calculated analytically and is given in the textbook 
of Jackson\cite{Jackson}:
\begin{equation}
\label{ana_photonflux}
f(x) = \frac{\alpha Z^2}{\pi} \, \frac{1}{x} \, \bigg[ 2 Y K_0(Y) K_1(Y) - 
Y^2 (K_1^2(Y) - K_0^2(Y)) \bigg] \; ,
\end{equation}
where $K_0$ and $K_1$ are modified Bessel functions and $Y= x M_A b_{min}$. 

The different photon spectra of high-energy protons discussed above are shown in 
Fig.~\ref{photon_spectrum} (left). Including the factor $(Q^2 - Q_{min}^2)/Q^2$ in the integral in 
Eq.~\ref{f_x} leads to a reduction in the photon flux compared with the result of Drees and 
Zeppenfeld. Including the effect of the magnetic moment of the proton (Kniehl) gives a   
photon flux somwhat higher than that of Eq.~\ref{JN} but lower than that of Drees and Zeppenfeld. 
The differences between the various approaches are largest for large values of $x$. At $x = 0.05$, 
Eq.~\ref{DZ} and \ref{JN} deviate from the result of Kniehl by about 15\%; at $x = 0.01$ the 
deviation is about 8\%. All three approaches give results broadly similar to that of a point charge 
with a minimum impact parameter of $b_{min} =$~0.7~fm (dash-dotted curve in the figure). 

It should be noted that the spectra above are for a single proton. In a proton-proton ($p+p$) collision, 
the effective photon flux may be 
reduced if one requires both protons to remain intact and not interact hadronically. This 
introduces an uncertainty in the effective photon spectrum. 

The photon spectra can be converted to an equivalent photon luminosity by multiplying 
$f(x)$ by the corresponding beam luminosity, ${\cal L}$. The result for Au+Au interactions at RHIC and 
Pb+Pb and $p+p$ interactions at the LHC are shown in Fig.~\ref{photon_spectrum} (right). The beam luminosities 
are taken from \cite{Eidelman:2004wy}. Comparing the photon 
luminosities in Pb+Pb and $p+p$, one sees that, at low photon energies, the luminosity in $p+p$ is higher than 
in Pb+Pb by a factor ${\cal L}_{pp}/(Z^2 {\cal L}_{PbPb})$. The photon spectrum, furthermore, extends to 
higher energies in $p+p$ collisions because of the higher beam energy ($\sqrt{s} =$~14 vs. 5.5 TeV at the LHC) 
and the larger momentum transfers. 

The photon luminosities in Fig.~\ref{photon_spectrum} (right) have been calculated from 
Eq.~\ref{ana_photonflux} with minimum impact parameters of $b_{min} =$ 0.7~fm (pp), 14.0~fm (AuAu), 
and 14.2~fm (PbPb). 
To be consistent with Ref.~\cite{Klein:2003vd}, the photon spectrum of Drees and Zeppenfeld 
(Eq.~\ref{DZ}) will be used for $p+p$ collisions in the rest of this paper.

\section{Particle Production Mechanisms}

In a collision between two nuclei or protons, the photons from one of the projectiles may interact 
with the other in a variety of ways. 
In the equivalent photon approximation, it is assumed that the total cross section factorizes into 
a photon flux and a two-photon or photonuclear cross section. The total two-photon cross section can then 
be written
\begin{equation}
\label{twophoton}
\sigma(A+A \rightarrow A+A+X) = \int_0^1 \int_0^1 f(x_1) f(x_2) \sigma_{\gamma \gamma}(\hat{s}) dx_1 dx_2 
\end{equation}
where the two-photon cross section, $\sigma_{\gamma \gamma}(\hat{s})$, is evaluated at a $\gamma \gamma$ 
center-of-mass energy squared $\hat{s} = x_1 x_2 s$. Here, $s$ is the collision center-of-mass energy 
squared and A can represent a nucleus or a proton (A=1). 

\begin{figure}
\begin{center}
\label{ppv}
\includegraphics[width=9cm]{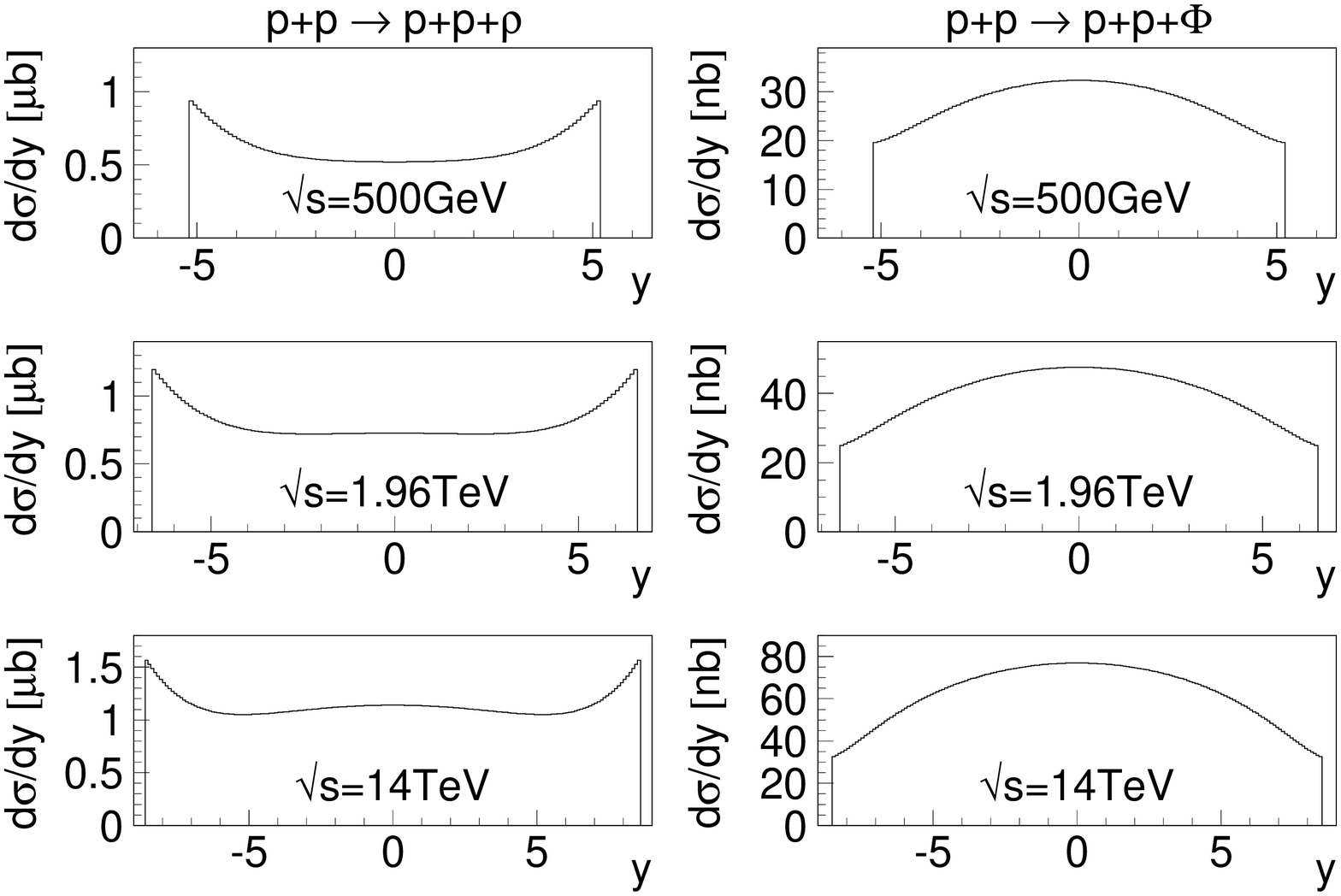}
\caption{Cross section for exclusive photoproduction of $\rho^0$ and $\phi$ in $p + p$ and $p + \overline{p}$ 
collisions at three different energies. See the text for a discussion of the sharp cut-off at the 
edges of the distributions.}
\end{center}
\end{figure}

Similarly, the cross section for photon-proton or photon-nucleus collisions can be written 
\begin{equation}
\sigma(A+A \rightarrow A+X) = \int_0^1 f(x) \sigma_{\gamma A}(\hat{s}) dx  
\end{equation}
where $\hat{s} = x s$. 

In a two-photon interaction, any charged particle/anti-particle pair can in principle be produced,
for example leptons $\gamma \gamma \rightarrow e^+e^-$, $\mu^+ \mu^-$; quarks 
$\gamma \gamma \rightarrow q \overline{q}$; or 
heavy vector bosons $\gamma \gamma \rightarrow W^+ W^-$. 
A $q \overline{q}$-pair can appear either as a bound state (meson) or fragment into two jets of 
hadrons. Two-photon interactions may be used to probe the internal structure of hadrons. 

The photon-hadron interactions can be divided into exclusive interactions, in which a certain particle 
or state is produced while the target remains in the ground state or is only internally excited, and inclusive 
interactions, in which the produced particle is accompanied by one or more others from the breakup of 
the target. The exclusive photon-induced interactions are dominated by exclusive vector meson production, 
$\gamma A \rightarrow VA$. Examples of inclusive interactions that have been studied are $q \overline{q}$ 
production through photon-gluon fusion\cite{Greiner:1994db} and photoproduction of jets 
through photon-parton interactions\cite{Vogt:2004yr}. 

\newpage

At high-energy accelerators, the production of $e^+e^-$-pairs have been studied in fixed target 
experiments with nuclear beams at the Bevalac (Berkeley)\cite{Belkacem:1993wq}, at the 
BNL Alternating Gradient Synchrotron\cite{Belkacem98}, and at the 
CERN SPS\cite{Baur:1994gv,Vane:1992ms}. At the heavy-ion collider RHIC, coherent photonuclear 
production of $\rho^0$ mesons\cite{Adler:2002sc} and two-photon production of 
$e^+e^-$-pairs\cite{Adams:2004rz} have been studied by the STAR collaboration. 
STAR has also published preliminary results on the interference between the two  $\rho^0$ 
production sources\cite{Klein:2004kq}. 
Recently, some 
first results on coherent $e^+e^-$-pair production at high invariant masses in Au+Au collisions at 
RHIC have been presented by the PHENIX collaboration\cite{Silvermyr04}. 

\begin{figure}
\begin{center}
\label{jpsi_coulomb}
\includegraphics[width=10cm]{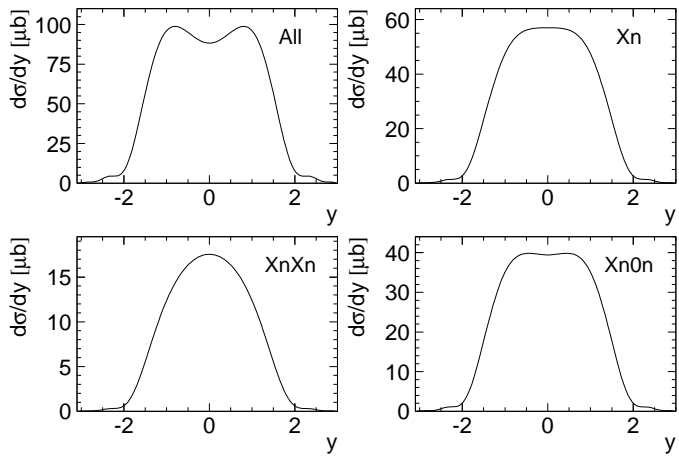}
\caption{Differential cross sections for $Au+Au \rightarrow Au+Au+ J / \Psi$ at $\sqrt{s} =$~0.2 TeV 
for different Coulomb breakup modes. ``All'' is the total coherent cross section. ``Xn'' corresponds 
to $J / \Psi$ production in coincidence with Coulomb breakup of at least one of the nuclei.  
This sample is divided into the cases where both nuclei break up (``XnXn'') and where 
only one of them breaks up (``Xn0n''). The total cross sections, integrated over all rapidities, 
are 290 $\mu$b (``All''), 159 $\mu$b (``Xn''), 115 $\mu$b (``Xn0n''), and 44 $\mu$b (``XnXn'').} 
\end{center}
\end{figure}

We will concentrate on exclusive photoproduction of vector mesons for the rest of this paper. We will, 
however, come back to two-photon production of lepton pairs in section 7. Internal nuclear excitations will be 
discussed further in section 5 in connection with multiple photon interactions in nucleus-nucleus collisions.

\section{Vector Meson Production in proton-proton Collisions}

The cross sections for exclusive photoproduction of heavy vector mesons in $p+p$ and $p + \overline{p}$ 
interactions have been calculated in Ref.~\cite{Klein:2003vd}. The calculations are based on the photon 
spectrum in Eq.~\ref{DZ} and parameterizations of the photon-proton cross sections derived from measurements at 
HERA and from QCD based models. 

Photoproduction of heavy vector mesons is a sensitive probe  
of the proton gluon distribution\cite{Ryskin:1992ui}, with $d \sigma /dt (t=0) \propto [g(x,M_V^2/4)]^2$.
For $J / \Psi$ and $\Upsilon$ production at mid-rapidity at the LHC, x-values of 
$\approx (2-7) \times 10^{-4}$ can be probed; at the Tevatron the corresponding numbers are 
$\approx (1.5 - 5) \times 10^{-3}$. 

Results on exclusive production of the lighter vector mesons $\rho^0$ and $\phi$ will be presented below. 
While these interactions are of less interest in themselves, they may constitute a significant background, 
particularly at the trigger level, to the production of heavier vector mesons. 
Furthermore, since the cross sections of $\gamma p \rightarrow \rho^0 p$ and 
$\gamma p \rightarrow \phi p$ are rather well known, 
measuring the exclusive $\rho^0$ and $\phi$ production in $p+p$ collisions may help to resolve the 
ambiguities in the photon spectrum mentioned in Section~2.

The rapidity distributions are shown in Fig.~2 and the total cross sections are listed in 
Table~\ref{table:1}. Table~\ref{table:1} also contains the total cross sections for $J / \Psi$ and
$\Upsilon(1S)$ mesons for comparison. 
The results have been obtained using the parameterizations 
of the photon-proton cross sections in Ref.~\cite{Klein:1999qj}. These show a discontinuity at 
the photon-proton threshold energy, $\sqrt{\hat{s}} = m_p + m_V$. While this discontinuity is not
expected to affect the total cross sections significantly, one should not expect $d \sigma/ dy$ 
to be correct near the edges. In nuclear collisions, this 
discontinuity is never a problem since the spectrum is cut-off by the nuclear form factor well above
the threshold. 

\begin{table}[htb]
\small
\caption{\small Cross sections for photoproduction of the vector mesons $\rho^0$, $\phi$, $J / \Psi$, and 
$\Upsilon(1S)$ in $p + p$ and $p + \overline{p}$ collisions.}
\label{table:1}
\newcommand{\m}{\hphantom{$-$}}
\newcommand{\cc}[1]{\multicolumn{1}{c}{#1}}
\renewcommand{\tabcolsep}{1.5pc} % enlarge column spacing
\renewcommand{\arraystretch}{1.2} % enlarge line spacing
\begin{center}
\begin{tabular}{@{}lcccc}
\hline
Accelerator                    & $\rho^0$ $[\mu b]$  & $\phi$ $[nb]$ & $J / \Psi$ $[nb]$  & $\Upsilon(1S)$ $[pb]$ \\ \hline
RHIC $\sqrt{s} =$ 0.2 TeV      & 4.4               & 180         & 3.0              & 2.3 \\
RHIC $\sqrt{s} =$ 0.5 TeV      & 6.4               & 290         & 7.0              & 12  \\
Tevatron $\sqrt{s} =$ 1.96 TeV & 10.6              & 520         & 23               & 120 \\
LHC   $\sqrt{s} =$ 14          & 19.6              & 1060        & 120              & 3500\\ \hline
\end{tabular}\\[2pt]
\end{center}
\end{table}

\section{Vector Meson Production and Multiple Photon Interactions in Nucleus-Nucleus Collisions}

The rates for exclusive vector meson production in ultra-peripheral nucleus-nucleus 
collisions at RHIC and the LHC are high. This was predicted\cite{Klein:1999qj}, and subsequent 
calculations\cite{Frankfurt:2001db} and experimental 
measurements\cite{Adler:2002sc} have confirmed this. 

Owing to the strong fields of heavy nuclei, it is possible to exchange more than one 
photon in the same event. The largest 
probabilities are obtained for multiple emission of low energy photons, which lead to an internal 
excitation of the target, in particular to excitations into a Giant Dipole Resonance (GDR). 
The GDR excitation is followed by disintegration of the nucleus, usually through emission of a 
single neutron. At heavy-ion colliders, these neutrons can be detected in 
forward ``Zero-Degree'' Calorimeters\cite{Baltz:1998ex,Chiu:2001ij}.

\begin{table}[htb]
\small
\caption{\small Comparison of the vector meson photoproduction and hadroproduction cross sections in 
$p+p$ collisions. The cross sections at the LHC are evaluated at the maximum nucleon-nucleon 
$\sqrt{s}$ in Pb+Pb collisions. 
The $J / \Psi$ hadroproduction cross section at $\sqrt{s} =$ 0.2 TeV has been 
measured by PHENIX\cite{Adler:2003qs}. The error is here the quadratic sum of the statistical, systematical, and 
absolute errors. The hadroproduction cross sections at 
$\sqrt{s} =$ 5.5 TeV are taken from predictions in the ALICE Physics Performance Report\cite{ALICE}.}
\label{table:2}
\newcommand{\m}{\hphantom{$-$}}
\newcommand{\cc}[1]{\multicolumn{1}{c}{#1}}
\renewcommand{\tabcolsep}{1.5pc} % enlarge column spacing
\renewcommand{\arraystretch}{1.2} % enlarge line spacing
\begin{center}
\begin{tabular}{@{}lcc}
\hline
Accelerator                    & Hadroproduction   & Photoproduction \\ 
                               & $p+p \rightarrow V + X$ & $p+p \rightarrow p+p+V $ \\ \hline
RHIC $\sqrt{s} =$ 0.2 TeV, $V = J / \Psi$     & 4.0$\pm$0.9 $\mu$b & 3.0 nb    \\ 
LHC $\sqrt{s} =$ 5.5 TeV,  $V = J / \Psi$     & 19-48 $\mu$b       & 54  nb     \\ 
LHC $\sqrt{s} =$ 5.5 TeV,  $V = \Upsilon(1S)$     & 190-280 nb         & 720 pb     \\ \hline
\end{tabular}\\[2pt]
\end{center}
\end{table}

It is possible to produce a vector meson in coincidence with Coulomb breakup of the nuclei. 
The first calculations of this process were done for vector meson production in coincidence with 
Coulomb breakup of both beam nuclei\cite{Baltz:2002pp}. This requirement coincided 
with the trigger used by 
STAR at RHIC at that time. Since then, the PHENIX experiment has introduced a 
trigger for ultra-peripheral collisions which requires a signal 
in only one of the Zero-Degree calorimeters\cite{Silvermyr04}. We present new calculations 
on $J / \Psi$ production 
in coincidence with single Coulomb breakup in Figure~3. The calculations are based on 
the model in\cite{Baltz:2002pp}, and the different Coulomb breakup modes are defined in 
the Figure caption. 

It is worth noting that for a heavy final state such as the $J / \Psi$, the requirement that at least 
one of the beam nuclei dissociates leads to a reduction in the production cross section of only 
about 45\%. For a light vector meson such as the $\rho$, the corresponding reduction is about a 
factor of 4.

\section{Photoproduction vs. Hadroproduction of Vector Mesons}

The coherent photoproduction cross sections of light vector mesons in nucleus-nucleus collisions 
are very large at RHIC and the LHC. The cross section for exclusive $\rho^0$ 
production, for example, is about 10\% of the total Au+Au cross section at RHIC, rising to 
about 50\% of the total Pb+Pb cross section at the LHC\cite{Klein:1999qj}. 

The photoproduction cross sections of heavy vector mesons in $p+p$ and A+A collisions are compared 
with the corresponding hadronic cross sections in Tables 2 and 3.  In $p+p$, the photoproduction 
cross sections are typically $\sim10^{-3}$ of the hadronic cross sections both at RHIC and the LHC. 
The same ratio applies for the $J/\Psi$ in Au+Au collisions at RHIC. At the LHC, the photoproduction 
cross sections in Pb+Pb collisions are typically a few percent of the hadroproduction cross sections. 
The hadronic nucleus-nucleus cross sections have been calculated assuming a $A^2$ scaling of the $p+p$ 
cross sections. The effects of nuclear gluon shadowing and the creation of a quark gluon plasma in 
central collisions have not been considered.

Although the photoproduction cross section is only a small fraction 
of the total hadronic cross section for vector mesons, separation of this reaction channel seems 
possible. 
The numbers in the tables provide an estimate of what rejection factors are needed.

\begin{table}[htb]
\small
\caption{\small Comparison of the vector meson photoproduction and hadroproduction cross sections in 
nucleus-nucleus collisions (Au at RHIC, Pb at the LHC). The total hadroproduction cross sections 
(integrated over all centralities) 
have been calculated as $A^2$ times the hadroproduction cross sections in $p+p$ collisions (Table 2). 
The photoproduction cross sections are from \cite{Klein:2003vd} and \cite{Klein:1999qj}.}
\label{table:3}
\newcommand{\m}{\hphantom{$-$}}
\newcommand{\cc}[1]{\multicolumn{1}{c}{#1}}
\renewcommand{\tabcolsep}{1.5pc} % enlarge column spacing
\renewcommand{\arraystretch}{1.2} % enlarge line spacing
\begin{center}
\begin{tabular}{@{}lcc}
\hline
Accelerator                    & Hadroproduction   & Photoproduction \\ 
                               & $A+A \rightarrow V + X$ & $A+A \rightarrow A+A+V $   \\ \hline
RHIC $\sqrt{s_{nn}} =$ 0.2 TeV, $V = J / \Psi$           & 160  mb           & 290 $\mu$b \\ 
LHC $\sqrt{s_{nn}} =$ 5.5 TeV,  $V = J / \Psi$           & 820-2100 mb       & 32  mb     \\ 
LHC $\sqrt{s_{nn}} =$ 5.5 TeV,  $V = \Upsilon(1S)$           & 8-12 mb           & 170 $\mu$b \\ \hline
\end{tabular}\\[2pt]
\end{center}
\end{table}

\section{Dielectron production, A+A $\rightarrow$ A+A+$e^+e^-$}

A $J / \Psi$ or $\Upsilon$ produced in an ultra-peripheral nucleus-nucleus collision may 
be detected through its decay into a $e^+e^-$ or $\mu^+\mu^-$ pair. We will focus on 
the dielectron channel here. A competing process with similar kinematics is  
the two-photon reaction $\gamma \gamma \rightarrow e^+e^-$. 
Both processes will lead to the production of an $e^+e^-$-pair with low transverse momentum and 
no other particles present in the same event. 

Pairs from the decay $J / \Psi \rightarrow e^+e^-$ will have an invariant mass 
determined by the mass (3.097 GeV) and natural width (91 keV) of the $J / \Psi$, whereas   
pairs from $\gamma \gamma \rightarrow e^+e^-$ will have a continuous distribution in 
$M_{inv}$. The distribution is peaked near threshold, $M_{inv} = 2m_e$, and falls 
rapidly with increasing 
$M_{inv}$; it is shown by the histogram in Fig.~4~a) for $M_{inv} >$~1.5~GeV. 

Using the Breit-Wheeler cross section $\sigma_{\gamma \gamma \rightarrow e^+ e^-}(\hat{s})$  
in Eq.~\ref{twophoton}, we find that the cross section for $M_{inv} > 1.5$~GeV is 1.4~mb. 
This is thus only about $4 \cdot 10^{-8}$ of the total $e^+e^-$ cross section of 33 kb in Au+Au collisions at 
$\sqrt{s_{nn}} =$~0.2~TeV\cite{Alscher:1996gn}. The $J / \Psi$ cross section multiplied by the 
branching ratio for $J / \Psi \rightarrow e^+e^-$ is 17~$\mu$b. The sum of the contribution from 
$\gamma \gamma \rightarrow e^+e^-$ and $J / \Psi \rightarrow e^+e^-$ is shown by the bars or 
crosses in Fig.~4. 

Fig.~4~b) shows the $M_{inv}$-distribution when cuts are applied on the emission angles of 
the $e^+$ and $e^-$. To simulate the coverage of a typical collider experiment, it is required that both 
particles are within $| \eta | <$~0.5, where $\eta = - \ln(\tan(\theta/2))$ is the pseudorapidity. 

The angular distribution from the decay of the $J / \Psi$ is given by \cite{Schilling:1969um}
\begin{equation}
\label{eq:vm}
\frac{dn}{d \cos(\theta)} \, \propto \, \sin^2(\theta) \; ,
\end{equation}
in the $e^+e^-$ center-of-mass system, while for the two-photon channel\cite{Brodsky:1971ud} 
\begin{equation}
\label{eq:epairs}
\frac{dn}{d \cos(\theta)} \, \propto \, 2 + 4 \left( 1 - \frac{4 m_e^2}{\hat{s}} \right) 
\frac{ \left( 1 - \frac{4 m_e^2}{\hat{s}} \right) \sin^2(\theta) \cos^2(\theta) + \frac{4 m_e^2}{\hat{s}}}{
\left( 1 - \left( 1 - \frac{4 m_e^2}{\hat{s}} \right) \cos^2(\theta) \right)^2} \; .
\end{equation}
Since the two-photon distribution is strongly peaked in the forward-backward direction,
the cut $| \eta | <$~0.5 removes more of the two-photon sample than of the vector meson sample. 

As long as only the natural width of the $J/\Psi$ is considered, the cross section for 
$J/\Psi \rightarrow e^+e^-$ is much larger than the continuum $e^+e^-$ cross section over 
the same range in $M_{inv}$. 
The experimentally observed $J/\Psi$ width (due to the finite detector resolution) is, however, 
always much larger than the natural width. The effect when a typical 
experimental $J/\Psi$ width of $\sim$110 MeV\cite{Adler:2003qs} is applied is shown in Fig.~4~c). 
The $J/\Psi$ peak can still be clearly identified, but the ratio of $J/\Psi$ to continuum is now 
much lower.

\begin{figure}
\begin{center}
\label{jpsi}
\includegraphics[width=9.0cm]{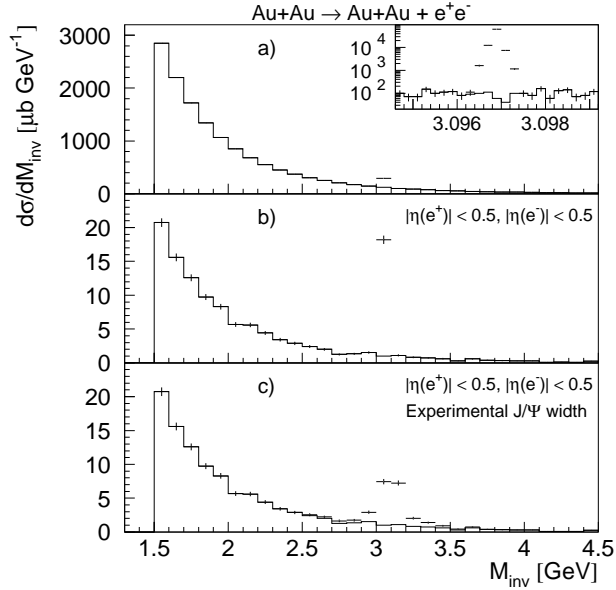}
\caption{The differential cross section $d \sigma /dM_{inv}$ for dielectron production in ultra-peripheral 
Au+Au collisions at $\sqrt{s_{nn}} =$~0.2~TeV. The histograms show the two-photon contribution, and the bars 
or crosses show the sum of the two-photon and $J / \Psi \rightarrow e^+e^-$ contribution. The inset in a) 
has an expanded $M_{inv}$ scale. The distributions have been calculated from a Monte Carlo simulation. 
700k $e^+e^-$-pairs with $M_{inv} >$~1.5~GeV have been generated, corresponding to an intergrated luminosity 
of 500 $\mu$b$^{-1}$.}
\end{center}
\end{figure}

The coherence requirement, which limits the $p_T$ of the produced vector meson to 
$\sqrt{2}/R \approx$50~MeV in 
Au+Au interactions, has been shown to provide very good background rejection. It can, however, not 
remove the contribution from $e^+e^-$-pair production through the two-photon channel, which is also 
peaked at very low total $p_T$. The two-photon $e^+e^-$ continuum may therefore 
constitute the main background to coherent $J/\Psi$ and $\Upsilon$ production in ultra-peripheral 
nucleus-nucleus collisions.

\section{Conclusions}

The spectrum of equivalent photons at current and future $p+p$ and heavy-ion colliders 
may be used to study photon-nucleon and photonuclear interactions at unprecedented energies.   
The exclusive photoproduction of heavy vector mesons has been considered here. The cross sections 
for photoproduction have been 
compared with the total hadronic cross sections. It has been shown that two-photon production of 
$e^+e^-$-pairs will be an important background to vector mesons produced in ultra-peripheral 
nucleus-nucleus collisions.

\end{document}